# Surface pressure impact on nitrogen-dominated USP super-Earth atmospheres

J. Chouqar,[1]★ J. Lustig-Yaeger,[2,3]★ Z. Benkhaldoun,[1] A. Szentgyorgyi,[4] A. Jabiri[1] and A. Soubkiou[1,5,6]

[1]*Oukaimeden Observatory, PHEA Laboratory, Cadi Ayyad University, BP 2390 Marrakech, Morocco*
[2]*Johns Hopkins University Applied Physics Laboratory, Laurel, MD 20723, USA*
[3]*NASA NExSS Virtual Planetary Laboratory, University of Washington, Box 351580, Seattle, WA 98195, USA*
[4]*Center for Astrophysics | Harvard & Smithsonian, 60 Garden Street, Cambridge, MA 02138, USA*
[5]*Departamento de Fisica e Astronomia, Faculdade de Ciencias, Universidade do Porto, Rua do Campo Alegre, P-4169-007 Porto, Portugal*
[6]*Instituto de Astrofisica e Ciencias do Espaco, Universidade do porto, CAUP, Rua das Estrelas, P-150-762 Porto, Portugal*

## ABSTRACT

In this paper, we compare the chemistry and the emission spectra of nitrogen-dominated cool, warm, and hot ultra-short-period (USP) super-Earth atmospheres in and out of chemical equilibrium at various surface pressure scenarios ranging from $10^{-1}$ to 10 bar. We link the one-dimensional VULCAN chemical kinetic code, in which thermochemical kinetic and vertical transport and photochemistry are taken into account, to the one-dimensional radiative transfer model, PETITRADTRANS, to predict the emission spectra of these planets. The radiative–convective temperature–pressure profiles were computed with the HELIOS code. Then, using PANDEXO noise simulator, we explore the observability of the differences produced by disequilibrium processes with the *JWST*. Our grids show how different surface pressures can significantly affect the temperature profiles, the atmospheric abundances, and consequently the emission spectra of these planets. We find that the divergences due to disequilibrium processes would be possible to observe in cooler planets by targeting HCN, $C_2H_4$, and CO, and in warmer planets by targeting $CH_4$ with HCN, using the NIRSpec and MIRI LRS *JWST* instruments. These species are also found to be sensitive indicators of the existence of surfaces on nitrogen-dominated USP super-Earths, providing information regarding the thickness of these atmospheres.

## 1 INTRODUCTION

The discovery of over 5000 exoplanets in the past two decades has unveiled a large and diverse population, far exceeding the diversity seen in our own Solar system. Among the distinct populations of small exoplanets, some of the most interesting are the so-called ultra-short-period (USP) exoplanets. These are planets that orbit at extremely short periods, from about 1 d down to less than 10 h (e.g. Howard et al. 2013; Dai et al. 2017), and even as short as ≈4 h, especially around M dwarfs (Ofir & Dreizler 2013; Rappaport et al. 2013; Smith et al. 2017). Planets in this group tend to be smaller than 2 $R_\oplus$, and appear to have bulk compositions similar to that of the Earth (Winn, Sanchis-Ojeda & Rappaport 2018). There is also evidence that some of them might have iron-rich compositions (Santerne et al. 2018). Their expected surface temperatures can reach thousands of Kelvin, due to the intense stellar radiation to which they are exposed, because of their proximity to the central star, allowing the detection of thermal emission from the planets' surfaces (Rouan et al. 2011; Demory et al. 2012; Sanchis-Ojeda et al. 2014). The induced stellar orbital velocities can be as high as a few ms$^{-1}$, allowing the planet masses to be measured with current technology even for stars as faint as $V = 12$ (Howard et al. 2013; Pepe et al. 2013). Although more than a hundred of these systems have been found by the *Kepler* mission, with which it was found that these exoplanets are extremely rare (about as rare as hot-Jupiters; Sanchis-Ojeda et al. 2014), only a handful of them have a precise radius and mass measurements, as the stars in the *Kepler* field are typically much too faint for precise radial velocity follow-up. Transit surveys like *TESS* (*Transiting Exoplanet Survey Satellite*), however, provide the perfect strategy to find these planets as they are designed to find short-period transiting exoplanets around bright stellar hosts, allowing us to explore the yet poorly understood property of mass and, thus, the bulk composition of these interesting extrasolar worlds.

A better understanding of this class of exoplanets will therefore require an increase in the sample of such planets that have accurate and precise masses and radii, which also includes estimates of the irradiation level and information about possible companions. A few such USP planets have measured masses and radii, which are consistent with a rocky composition (e.g. 55 Cnc e, Dawson & Fabrycky 2010; Demory et al. 2011; CoRoT-7b, Queloz et al. 2009; Kepler-10b, Batalha et al. 2011). The atmospheres of these small hot planets are poorly understood, so little is known about their composition, chemistry, and condensates. Several studies have been carried out to explore the compositions of these hot–rocky super-Earths. Schaefer, Lodders & Fegley (2012) and Schaefer & Fegley (2009) discuss the possible atmospheres of Corot-7b-like planets. The results of their modelling suggested that the atmospheres are composed primarily of Na, $O_2$, O, and SiO gas. Miguel et al. (2011) explored the composition of initial planetary atmospheres of close-in hot super-Earths in the *Kepler* planet candidate sample. The resultant atmospheric compositions are similar to those from Schaefer & Fegley (2009): the major constituents are Na, O, $O_2$, and SiO. Demory et al. (2016), Angelo & Hu (2017), and Bourrier et al. (2018), using interior composition models, found that the measured mass and radius of 55 Cnc e are consistent with a planetary scenario with a massive, high-mean-molecular-weight atmosphere. Angelo & Hu (2017) suggest a CO- or $N_2$-dominated envelope as plausible scenarios for the planet 55 Cnc e. Furthermore, there have been searches for molecular features (e.g. Na and HCN) in 55 Cnc e, using transit spectroscopy in the infrared (Ridden-Harper et al. 2016; Tsiaras et al. 2016) and high-resolution optical spectroscopy (Esteves et al. 2017), but the results are inconclusive. Miguel (2018) explored the possibility of a nitrogen-based atmosphere, for 55 Cnc e, using equilibrium chemistry, and found that the atmosphere might show spectral features of $NH_3$, HCN, and CO. Zilinskas et al. (2020a) adopted a chemical kinetics model to explore the possible composition of a nitrogen-dominated atmosphere and the observability of their spectral features with the future mission, for the benchmark 55 Cnc e. In a followed study, Zilinskas et al. (2020b) investigated the effects caused by thermal inversions on the chemistry and emission spectra of USP super-Earth nitrogen-dominated atmospheres.

★ E-mail: jamila.chouqar@ced.uca.ma (JC); jlustigy@uw.edu (JL-Y)

Since its science operations began in July 2018, NASA's *TESS* (Ricker 2015) has significantly expanded the size of this sample. An additional 24 planets with $R_p < 2\,R_\oplus$, $P < 1$ d, and host stars amenable to *JWST* observations have been discovered, with 13 of them having an accurate determination of mass and radius.[1] Our ability to characterize short-period terrestrial planets will improve drastically with the *JWST*, which will allow for the characterization of exoplanet atmospheres and surface properties via transmission and emission spectroscopy (Greene et al. 2016). These planets are among the Cycle 1 observations and will continue to be observed in Cycle 2 and beyond. However, we currently lack a reliable set of models to guide the planning of such observations and help their interpretation.

Nitrogen-dominated atmospheres of USP super-Earth planets have been widely studied with the benchmark 55 Cnc e. Effects of orbital distances and compositions with different abundances of hydrogen, C/O, and N/O ratios have been explored (Zilinskas et al. 2020b). A broader exploration as a function of surface pressure is also required. Surface pressure variations affect the thermal profile of the planet, hence the chemistry and the spectra. In this paper, we present a grid of thermal profiles, photochemical mixing ratios, and emission spectra for different surface pressure scenarios for cool (LP791-18b), warm (GJ-1252b), and hot (HD-213885b) USP super-Earths in order to explore the effects of surfaces and what to expect in current and upcoming observations. For each planet, we consider three scenarios that cover three different surface pressures ranging from $10^{-1}$ to 10 bar. These scenarios are chosen in light of previous studies on USP super-Earths. So far, only four USP super-Earths, 55 Cnc e (Angelo & Hu 2017), GJ-1252b (Crossfield et al. 2022), LHS-3844b (Kreidberg et al. 2019), and K2-141b (Zieba et al. 2022), have had their thermal infrared radiation measured. *Spitzer* 4.5 μm observations of these planets' eclipse and phase curves revealed no phase offset and suggested an upper limit to the atmospheric surface pressure. For example, the data set for LHS-3844b indicates $P_{sur} \leq 10$ bar (Kreidberg et al. 2019; Whittaker et al. 2022). Crossfield et al. (2022) compared the measurement of GJ-1252b to a large suite of atmospheric models and concluded that any atmosphere on GJ 1252b likely has a surface pressure of ≤10 bar. Furthermore, Angelo & Hu (2017) have shown that the 1.4 bar of photospheric pressure is required to explain the planet's heat redistribution efficiency value constrained by the phase curve of 55 Cnc e.

This paper is organized as follows: In Section 2, we outline the details of the model grid and the tools used to generate the radiative transfer models. In Section 3, we present our main findings. In Section 4, we assess the observability with *JWST*, determining to what extent *JWST* will help distinct disequilibrium features. Finally, we summarize our work in Section 5.

## 2 ATMOSPHERIC MODEL

### 2.1 Planet properties

For this work, we select three confirmed *TESS* USP planetary candidates, with a radius smaller than 2 $R_\oplus$ (see Table 1). GJ 1252b and HD 213885b have measured masses, while the mass of LP791-18b is estimated using Forecaster.[2] In Fig. 1, we plot the masses and radii of all the known USP planets, with radius <2 $R_\oplus$, having these parameters measured with a precision better than 25 per cent. We compare the masses and radii of our targets with the theoretical and empirical mass–radius relationship of Zeng, Sasselov & Jacobsen (2016), and we find that their compositions are consistent with Earth-like/rocky planets.

### 2.2 Atmospheric pressure–temperature profiles

We used the HELIOS[3] code to compute the temperature structure of atmospheres using self-consistent radiative–convective iteration (Malik et al. 2017). HELIOS requires fundamental parameters for both the star and the planet, as well as chemical composition. Estimates for those parameters were taken from the published articles (see Table 1). The opacity k-tables are constructed using pre-calculated opacities obtained from the opacity data base[4] (Grimm & Heng 2015). In this work, we consider an $N_2$-dominated atmosphere assuming a Titan-like composition (Miguel 2018), where the mole fractions of each element are N = 0.962 87, H = 0.03, C = 0.007, and O = 2.5 × $10^{-5}$. As absorbers, we include $H_2O$, $CO_2$, CO, HCN, $CH_4$, CN, $NH_3$, NH, $CH_3$, CH, $C_2H_4$, $C_2H_2$, NO, and OH. The chemical equilibrium mixing ratios are calculated with FASTCHEM[5] (Stock et al. 2018). FASTCHEM is tested between 100 and 6000 K to accurately calculate the thermochemical equilibrium mixing ratios from a chemical network of 550 gas-phase species, including ions. For the k-table sampling wavelength resolution, we use $\lambda/\Delta\lambda = 1000$, accounting for a range between 0.3 and 100 μm.

The ultraviolet (UV) spectrum of a planet's host star is a critical input to chemical kinetics models. The UV environment of the star dominates the photochemistry and therefore the resulting

---

[1] https://exoplanetarchive.ipac.caltech.edu/

[2] https://github.com/chenjj2/forecaster
[3] https://github.com/exoclime/HELIOS
[4] https://chaldene.unibe.ch/data/Opacity3/
[5] https://github.com/exoclime/FastChem

**Table 1.** Relevant stellar and planetary properties.

|  | $R\,(R_*)$ | $M\,(M_*)$ | $T_{\rm eff}$ (K) | $\log g_{\rm s}$ (ms$^{-2}$) | [M/H] |  |  |  |
|---|---|---|---|---|---|---|---|---|
| LP791-18[a] | 0.171 | 0.139 | 2960 | 5.11 | −0.09 | – | – | – |
| GJ1252[b] | 0.391 | 0.381 | 3458 | 4.83 | 0.1 | – | – | – |
| HD213885[c] | 1.1011 | 1.068 | 5978 | 4.38 | −0.04 | – | – | – |
|  | $R\,(R_\oplus)$ | $M\,(M_\oplus)$ | $g$ (ms$^{-2}$) | $T_{\rm eq}$ (K) | Period (d) | $f^{\rm d}$ | $f^{\rm e}$ | $f^{\rm f}$ |
| LP791-18b[a] | 1.12 | 1.5[g] | 14 | 599 | 0.95 | 0.60 | 0.39 | 0.27 |
| GJ-1252b[b] | 1.19 | 2.09 | 14.4 | 1089 | 0.52 | 0.63 | 0.48 | 0.29 |
| HD-213885b[c] | 1.74 | 8.83 | 28.5 | 2130 | 1.00 | 0.66 | 0.61 | 0.39 |

*Note.* The equilibrium temperatures are calculated assuming zero Bond albedo and complete heat circulation between the day and night hemispheres.
[a]Crossfield et al. (2019).
[b]Shporer et al. (2020).
[c]Espinoza et al. (2019).
[d]Heat redistribution factor, for 0.1 bar surface pressure.
[e]Heat redistribution factor, for 1 bar surface pressure.
[f]Heat redistribution factor, for 10 bar surface pressure.
[g]Masses are estimated based on Chen & Kipping (2016).

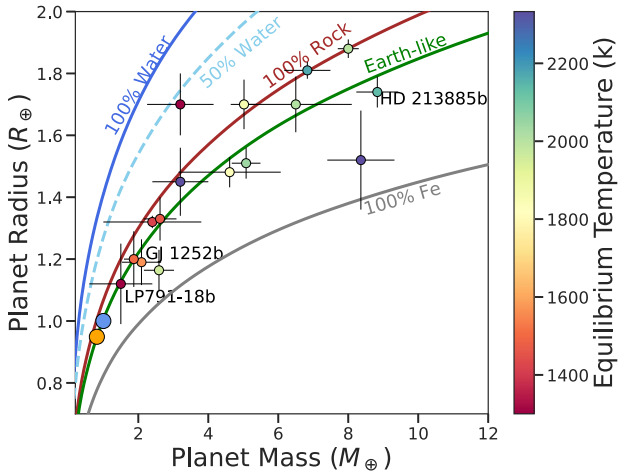

**Figure 1.** Mass–radius for USP planets, with mass–radius measurements better than 25 per cent from http://www.astro.keele.ac.uk/jkt/tepcat/ and for our USP *TESS* candidates (see Table 1), colour coded by their equilibrium temperature. Two-layer models from Zeng et al. (2016) are displayed with different lines and colours. 'Earth-like' here means a composition of 30 per cent Fe and 70 per cent MgSiO$_3$, whereas '100 per cent Rock' means a composition of 100 per cent MgSiO$_3$. Earth and Venus are identified in this plot as pale blue and orange circles, respectively.

atmospheric constituents (i.e. the rates of photolysis reactions are determined by the UV flux). Unfortunately, the UV spectra of many exoplanet hosts have not been measured, making it difficult to accurately model the photochemistry that occurs in their planets' atmospheres. However, for cases in which the host UV spectrum is unavailable, a reconstructed or proxy spectrum will need to be used in its place. For each of our M dwarf stars, we scale the UV flux from AD Leo's observed UV spectrum[6] to the other spectral types according to Rugheimer et al. (2015). We then combine the synthetic PHOENIX spectrum with the scaled observations for AD Leo. For the hotter star HD-213885, we scale the shortwave VPL solar spectrum,[7] in combination with PHOENIX models. The stellar spectra are taken from the PHOENIX online library and interpolated, using the PYSYNPHOT[8] package, for the stellar temperatures, $\log g$ star, and [M/H] star (see Table 1).

The efficiency of the horizontal heat transport is parametrized by the heat redistribution factor in a 1D radiative transfer model (e.g. Malik et al. 2019). To estimate the heat redistribution factor, we use the scaling theory of Koll (2022) that depends on the longwave optical depth of the atmosphere. The calculation of the heat redistribution factor (*f*) is done through an iterative process, as it is dependent on the optical depth of the planet's atmosphere, which is, in turn, dependent on the atmospheric properties of the planet. The optical depth is, thus, determined self-consistently with the Temperature Pressure (TP) profile. Additionally, the choice of *f* remains self-consistent with the bulk atmospheric composition because the iterative optical depth calculation takes the opacity sources into effect. For the surface, we use a spectrally uniform albedo of 0.01. This assumption is supported by observations (Rowe et al. 2006; Rogers et al. 2009; Crossfield et al. 2019) and modelling (Sudarsky, Burrows & Pinto 2000; Burrows, Ibgui & Hubeny 2008; Miller-Ricci & Fortney 2010) of hot exoplanets, which found very low albedos. Table 1 shows values of the heat redistribution factor (*f*) in our simulations with 0.1, 1, and 10 bar surface pressure. Atmospheres with a 10 bar surface pressure have a heat redistribution that clearly deviates from bare rock, whereas in thinner atmospheres (0.1 bar surface case) heat redistribution becomes inefficient. Unlike LP791-18b and GJ-1225b, the scaling predicts that the HD-213885b planet requires an atmosphere thicker than 1 bar surface pressure to exhibit significant heat redistribution as the heat redistribution is less efficient on hotter planets (Koll 2022). The dayside-average temperature profiles for the different surface pressure scenarios (0.1, 1, and 10 bar) are shown in Fig. 2.

### 2.3 Chemistry model

There are two primary disequilibrium processes that modify abundances within exoplanetary atmospheres. The first one is photochemistry that includes the dissociation and ionization of atmospheric constituents by stellar radiation. The second is transport-induced quenching that refers to the mechanism by which the atmospheric composition is driven away from chemical equilibrium as a result of the dominance of transport processes like molecular and eddy diffusion. In this paper, we use the open-source photochemical kinetic code VULCAN[9] (Tsai et al. 2017, 2021b) to model the chemistry occurring in our USP super-Earth atmospheres. VULCAN solves a set of continuity equations for different molecular species. We use the N–C–H–O photochemical network with about 700 reactions and over 50 molecular species. The network is available as part of the VULCAN package. VULCAN is implemented with the equilibrium chemistry code FASTCHEM (Stock et al. 2018) to initialize a state in chemical equilibrium (see Tsai et al. 2017, for a more detailed description of the model). Physical processes, such as photochemistry, and transport-induced quenching are included. Vertical transport of gases in the atmosphere is set for 1D models by the so-called eddy diffusion coefficient $K_{zz}$. As in Zilinskas et al. (2020b), we assumed a commonly used value for rocky planets of $K_{zz} = 10^8$ cm$^2$ s$^{-1}$. The molecular diffusion coefficients were computed following Banks & Kockarts (1973) and Moses et al. (2000) and they depend on temperature, total number density, and mass of the diffusing species. We calculate the molecular diffusion coefficients for each species according to the dominant atmospheric constituent, which in our case is molecular nitrogen. The computed chemical abundances are shown in Fig. 4.

---

[6]http://depts.washington.edu/naivpl/content/spectral-databases-and-tools
[7]http://depts.washington.edu/naivpl/content/spectral-databases-and-tools
[8]https://pysynphot.readthedocs.io/en/latest/index.html

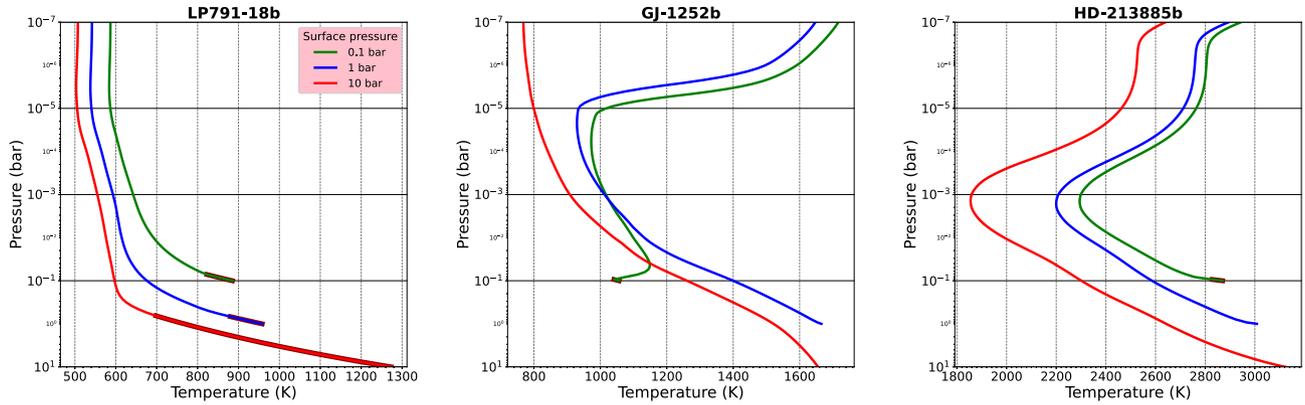

**Figure 2.** Temperature–pressure profiles in radiative–convective equilibrium computed using HELIOS for each of the considered planets in an $N_2$-dominated atmospheres. Each column corresponds to a different planet. For each of the cases, the profiles are shown as a function of surface pressure. A convective zone (orange) appears in the lower regions of some planets.

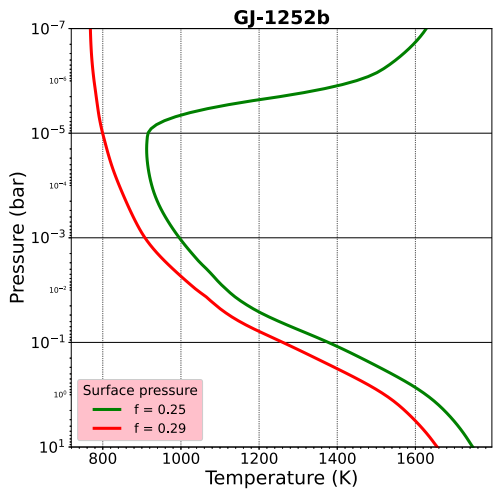

**Figure 3.** Temperature–pressure profiles of GJ-1252b planet assuming the 10 bar surface pressure. Colours correspond to different heat redistribution factors, which were calculated using the formulation of heat redistribution for rocky planets (Koll 2022). A heat redistribution factor of 0.29 tends to cool the atmosphere by about 100 K.

### 2.4 Synthetic spectra

In order to predict emission spectra of these planets, we use the PETITRADTRANS[10] radiative transfer and retrieval code (Mollière et al. 2019). We use the low-resolution option ($\lambda/\Delta\lambda = 1000$) of PETITRADTRANS that follows the correlated-$k$ approximation. The species that we consider to generate the synthetic spectra are $H_2$, $H_2O$, $CO_2$, CO, HCN, $CH_4$, CN, $NH_3$, NH, $CH_3$, CH, $C_2H_4$, $C_2H_2$, NO, and OH. We also consider Rayleigh scattering for $N_2$, $H_2$, and He, as well as collision-induced absorption for the $N_2$–$N_2$, $H_2$–$H_2$, and $H_2$–He pairs. The line opacities of most of these species can be found in table 2 of Mollière et al. (2019). The opacities of $C_2H_4$, $CH_3$, CH, CN, NH, and NO were taken from the Exomol data base[11]; these are already in the PETITRADTRANS format and can be used in a plug-and-play fashion. The opacities are available only for the low-resolution mode of PETITRADTRANS ($\lambda/\Delta\lambda = 1000$); for more details, we refer the reader to see the code documentation page[12] and Chubb et al. (2020). The molecular abundances computed by VULCAN in one dimension were used as inputs for PETITRADTRANS. The resulting emission spectra are shown in Fig. 5.

## 3 MODELLING RESULTS

### 3.1 Temperature profiles

HD-213885b is the hottest planet modelled in this paper. It has an equilibrium temperature of 2130 K (adopting a zero albedo) and its thermal profiles for each surface pressure are shown in Fig. 2. Stratospheric inversions on this planet are common for all surface pressures. The inversions found here are different from those on hot Jupiters since we do not include other absorbers such as TiO and VO.

---

[9]https://github.com/exoclime/VULCAN
[10]http://gitlab.com/mauricemolli/petitRADTRANS
[11]http://www.exomol.com/
[12]https://petitradtrans.readthedocs.io/en/latest/

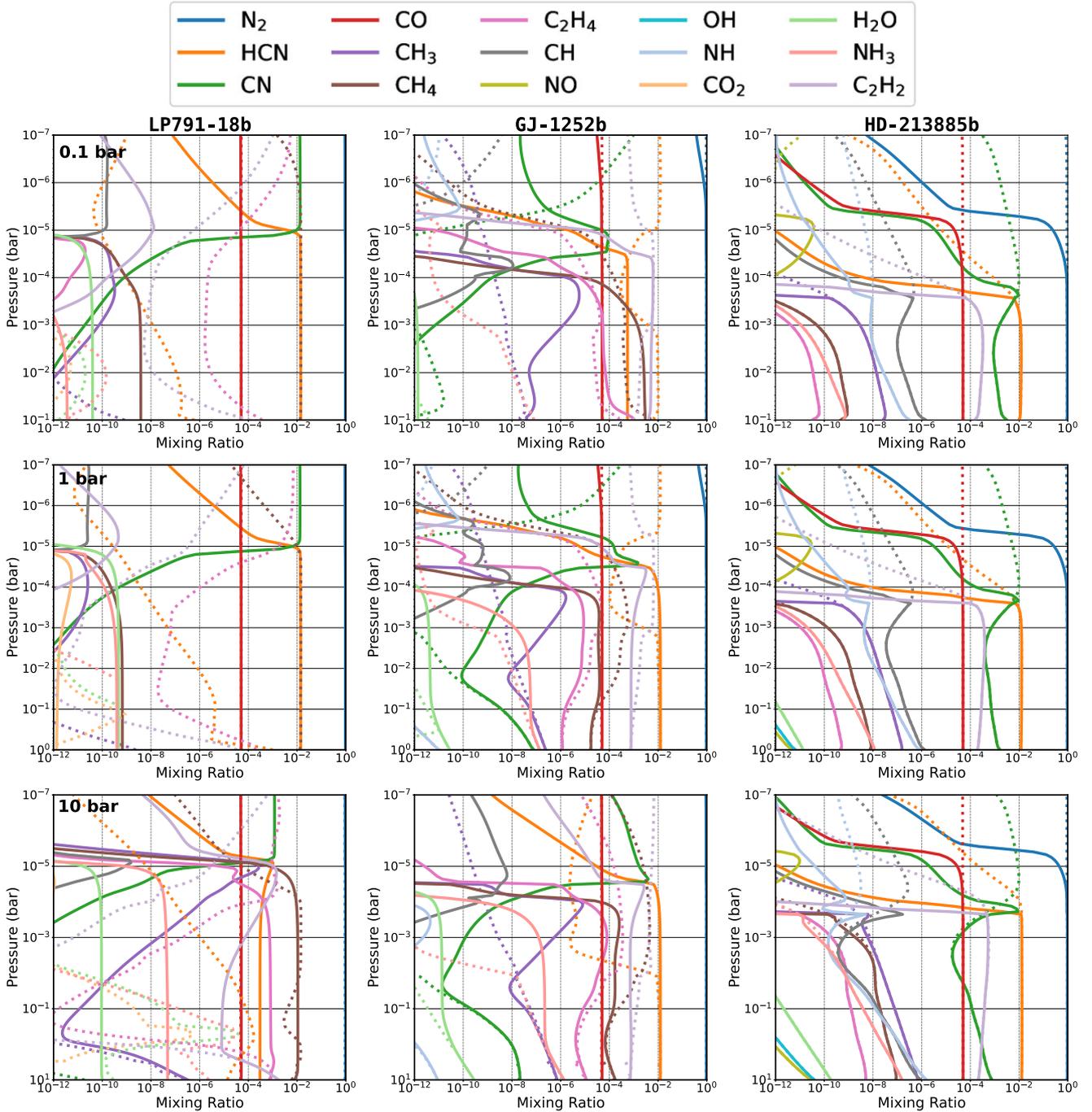

**Figure 4.** VMRs of major chemical species in a nitrogen-dominated atmosphere for each of the USP super-Earths considered in this work. Each row represents a different surface pressure. The VMRs are calculated using chemical kinetic with radiative–convective temperature profiles (see Fig. 2). The dotted profiles are the thermochemical equilibrium abundances. The solid profiles are the disequilibrium chemical abundances calculated with vertical mixing ($K_{zz} = 10^8$ cm$^2$ s$^{-1}$) and photochemistry.

Instead, the inversions are caused by strong atmospheric absorption due to the presence of the CN molecule, which is known to be a significant shortwave absorber (Zilinskas et al. 2020a). The increase in CN abundance causes strong temperature inversions above $10^{-3}$ bar (Fig. 2). A thicker atmosphere with a surface pressure of 10 bar can transport heat more efficiently with a heat redistribution factor ($f$) reaching up to 0.39 (Table 1). Although not fully achieving global heat redistribution ($f = 0.25$), this significantly impacts the energy budget, reducing temperatures by about 450 K compared to a thin atmosphere with 0.1 bar surface pressure. The atmosphere in the latter case is expected to be inefficient in transporting heat, confining it to the dayside of the planet ($f = 0.66$).

GJ-1252b thermal profiles are shown in Fig. 2. This planet has an equilibrium temperature of 1089 K (assuming a zero albedo). Similar to HD-213885b, the formation of CN in the upper atmosphere of GJ-1252b, due to the thermal dissociation of many molecules (e.g. $NH_3$, $CH_4$, and $C_2H_4$), causes temperature inversions above $10^{-5}$ bar for only simulations with 0.1 and 1 bar surface pressures. The 10 bar surface pressure case for GJ-1252b differs. As can be seen from Fig. 3, the atmosphere is effectively cooled by the $f$ convergence (heat redistribution factor of 0.29) by about 100 K, which prevents inversions. Here, CN appears to have a significant temperature dependence. Above a certain threshold, we have CN actively absorbing and below this temperature, CN appears to be ineffective (Fig. 3). This also explains why the inversion occurs in the atmospheres with lower surface pressure since those atmospheres are warmer than the 10 bar atmosphere. This effect is due to the temperature dependence of the mixing ratio of CN.

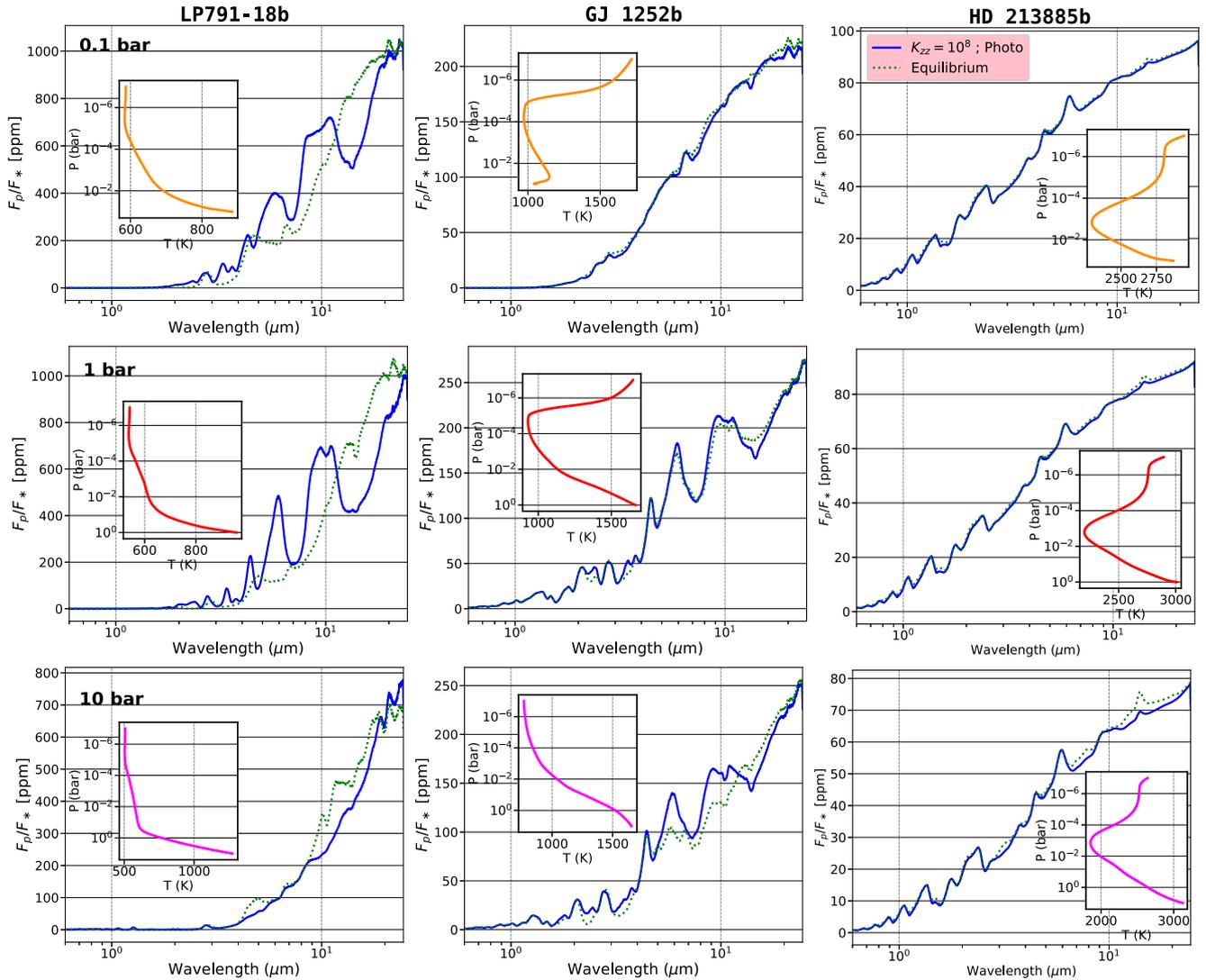

**Figure 5.** Effect of photochemical kinetic and vertical transport on the synthetic emission spectra of HD-213885b, GJ-1252b, and LP791-18b. Each row represents a different surface pressure. Each column corresponds to a different planet. Inlaid: temperature–pressure profiles. The dotted profiles are the thermochemical equilibrium abundances. The solid profiles are the disequilibrium chemical abundances calculated with vertical mixing ($K_{zz} = 10^8$ cm$^2$ s$^{-1}$) and photochemistry. The models shown are smoothed for clarity.

Moving now to our coldest case; LP791-18b planet. At this temperature, methane becomes the major opacity source. The presence of methane (greenhouse absorber) increases the temperatures with increasing pressure (Fig. 2). At pressure levels where the atmosphere is optically thick, radiative cooling is inhibited and these layers warm until the higher temperatures establish energy balance again (through increased blackbody emission).

As for the variation with surface pressure, the heat transport scaling could also have a significant impact. Our model predicts that as surface pressure increases, there is less overall energy available to heat the dayside atmosphere. When a planet's atmosphere is thicker (e.g. 10 bar case), it is more efficient at transporting heat from the dayside to the nightside (e.g. $f = 0.27$; see Table 1), which can lower the overall temperature of the dayside atmosphere. Consequently, this cooling effect can extend to the upper atmosphere, leading to lower temperatures at higher surface pressures. This, in turn, leads to an increase in the temperature difference between the bottom of the atmosphere and the top of the atmosphere.

We note that we ignore the effect of chemical disequilibrium on the temperature structure of planetary atmospheres. Non-equilibrium processes (transport and photochemistry) have an influence on the chemical composition and thereby the temperature (Drummond et al.

2016). This temperature change can have important consequences for the chemical abundances themselves as well as for the simulated emission spectra and should be taken into account in future works.

## 3.2 Equilibrium versus disequilibrium

In this section, the effects of the three main chemical processes that control and modify the atmospheric composition of exoplanets are compared. Thermochemical equilibrium, transport-induced quenching, and photochemistry can all combine to play a significant role in influencing atmospheric properties and thus should be considered when investigating the chemistry of these exoplanets. Here, we explore how disequilibrium processes can affect the predicted composition and spectra at different surface pressures. We run two model scenarios: equilibrium (i.e. no-diffusion, no-photochemistry) and disequilibrium (i.e. diffusion and photochemistry) with surfaces located at three different pressure levels (0.1, 1, and 10 bar). The resulting chemical abundances and thermal spectra of our hot super-Earth $N_2$-dominated atmospheres with varying surface pressure are shown in Figs 4 and 5, respectively. Both the chemistry and the predicted spectra of our hot-super-Earth exoplanets depend strongly on the adopted thermal structure. Different surface levels with different surface pressures and temperatures are shown to have significant impacts on the final chemical make-up and spectra of the atmospheres (e.g. see Fig. 6). However, we note that CO abundance is weakly affected by the changes in surface pressures, hence temperatures. The amount of CO is essentially the same in all three planet cases. Temperature inversions have no strong effect on its abundance, except for lowest pressure regions, where it is increasingly dissociated with hot inverted atmospheres. This is consistent with Zilinskas et al. (2020b) findings.

### 3.2.1 HD-213885b case

The equilibrium and disequilibrium emission spectra of HD-213885b planet look relatively similar for all the surface pressure scenarios. They are all dominated by the spectral features of HCN and CN. The 0.1 and 1 bar surface results (2880 K $\leq T_{surf} \leq$ 3000 K) produce similar abundances of HCN and CN (see Fig. 4). Under these surface pressure and temperature conditions, the disequilibrium spectra look just like the equilibrium ones (Fig. 5). For the hotter 10 bar surface case ($T_{surf} >$ 3100 K), the disequilibrium model starts to deviate from the equilibrium model, exhibiting decreased volume mixing ratios (VMRs) of HCN. The thermochemical equilibrium profile of HCN has a higher VMR in the observable part of the atmospheres ($10^{-4}$–1 bar). When vertical transport and photochemistry are included in the model, the HCN VMR for the 10 bar surface case decreases from its expected equilibrium value by more than three orders of magnitude in the upper portion of the observable region of the atmosphere, as a result of photochemistry. This reduces the emitted flux at around 3.2, 8, and 13.9 μm wavelengths as can be seen in Fig. 5.

### 3.2.2 GJ-1252b case

For low surface pressure ($P_{surf}$ = 0.1 bar and $T_{surf} \approx$ 1050 K), methane dominates both equilibrium and disequilibrium spectra with a contribution from HCN. Compared to the equilibrium model, the disequilibrium model predicts minimal amounts of HCN by more than an order of magnitude, as a consequence of photochemistry

(Fig. 4), giving rise to an absorption signature at 13.9 μm. For the 1 bar case, the surface temperature becomes hot ($T_{surf} \approx$ 1670 K) enough that HCN becomes the dominant active species in the atmosphere. The methane abundance (Fig. 4) drops and shows minor contributions in the thermal spectra between 2 and 4 μm that are deeper in the equilibrium model by about 15 ppm compared to the disequilibrium one. The HCN abundance in the upper observable regions ($10^{-3}$–$10^{-4}$) is increased by a factor of $10^2$ when chemical disequilibrium is taken into account, leading to an increase in the HCN absorption band depth at 13.9 μm by about 20 ppm, with respect to the equilibrium depth (Fig. 5). Further, the predicted emission spectra of GJ-1252b are unaffected by the thermal inversions as they occur above the $10^{-5}$ bar level. The lower pressures do not contribute significantly to the thermal spectra, and consequently, the photochemically produced CN. For the 10 bar surface case, the surface temperature is only ~10 K lower compared to the 1 bar surface case; thus, most species have similar VMR profiles in the observable regions for both equilibrium and disequilibrium models. Only HCN, $CH_4$, and $C_2H_4$ species show notable sensitivity to this temperature differential. Methane and ethylene abundances are enhanced for both models. Thermochemistry reduces the HCN abundance by a factor of $10^{-1}$ with respect to the 1 bar case. Thus, large differences between equilibrium and disequilibrium spectra can be seen in Fig. 5. The overall shape of the disequilibrium spectrum is due to HCN with methane and ethylene contributing between 2–4 and 9–13 μm, respectively. Unlike the disequilibrium spectrum, the shape of the equilibrium emission is determined by $CH_4$ with HCN and $C_2H_4$ showing absorption at 5–7 and 11 μm, respectively.

Furthermore, Fig. 6 is included to show the sensitivity of disequilibrium models to different surface pressures. The 0.1 bar surface pressure case displays a notable departure from the remaining cases (1 and 10 bar). This is due to the predominance of $CH_4$ in the 0.1 bar surface pressure atmosphere, in contrast to the prevalence of HCN in the higher surface pressure atmospheres.

### 3.2.3 LP791-18b case

In the cool 0.1 bar surface atmosphere ($T_{surf} \approx$ 890 K), the thermochemical equilibrium model is dominated by methane. The thermal emission flux from 9 to 13 μm is reduced by about 100 ppm due to the significant contribution of $C_2H_4$. Also, $C_2H_2$ exhibits an absorption near 13.9 μm. With photochemistry and transport, the mixing ratio of HCN is considerably enhanced by 3–6 orders of magnitude in the observable atmosphere, whereas $CH_4$, $C_2H_2$, and $C_2H_4$ are depleted over equilibrium expectations. Thus, the overall shape of the disequilibrium emission model is determined by HCN features. The HCN enhancement is mostly caused by the coupled $NH_3$–$CH_4$ photochemistry, which is initiated by reactions of the form $CH_3$ + $NH_2$ + M → $CH_3NH_2$ + M, where M is any third atmospheric molecule or atom, followed by reactions with atomic H that lead to HCN formation (Moses et al. 2010, 2011; Tsai et al. 2017). In the 1 bar surface pressure scenario, methane and ethylene show the same emission features in the thermochemical model as in the 0.1 bar surface case. The 1 bar surface temperature ($T_{surf} \approx$ 970 K) is warm enough to deplete $C_2H_2$ and enhance the HCN abundance, leading to a remarkable absorption feature at around 13.9 μm. When disequilibrium processes are taken into account, the 1 bar surface model produces a similar abundance of HCN compared to the 0.1 bar surface model, resulting in a similar emission spectrum with differences in feature widths and depths. In the 10 bar surface atmosphere, the surface temperature (>1290 K) is hot enough for thermochemistry to be effective in the deep layers; thus, methane becomes the dominant species. Along with methane, the equilibrium model shows additional contributions from different species like HCN at 13.9 μm, $NH_3$ at 10 μm, and $NH_3/H_2O$ feature at around 20 μm. The disequilibrium model exhibits different features in the emission spectrum, mainly due to methane and ethylene that enable it to be distinguished from the equilibrium model. Also, we see that the $C_2H_4$/CO feature shows a decrease in the emission signal at 4.5 μm.

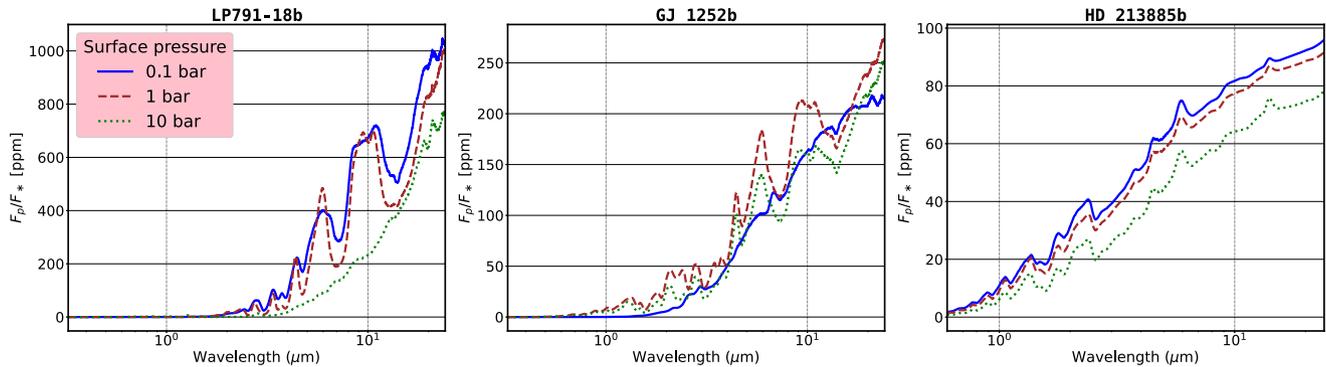

**Figure 6.** Sensitivity of disequilibrium spectra to different surface pressures for each planet.

Generally, in the 0.1 and 1 bar cases, the surface level is located well above the quenching point for most species, leading atmospheric compositions to deviate significantly from the equilibrium, as opposed to being tied to an equilibrium abundance at depth. As a result, the VMRs of photochemically fragile species ($CH_4, C_2H_2, C_2H_4$, and $NH_3$) drop significantly compared to the 10 bar surface case, while the VMRs of the more photochemically stable species (CO, HCN) increase as they become the major end products for the oxygen, carbon, and nitrogen in the observable atmosphere (see Fig. 4). The 10 bar surface is also above the original quench point for numerous species such as HCN, $NH_3$, and $H_2O$, but the surface temperature is high enough for thermochemistry to fully recycle the hydrocarbon photochemical products back to methane and transport $CH_4$ back to the upper atmosphere.

Moreover, the disequilibrium models of thinner atmospheres (0.1 and 1 bar surface pressure cases) are primarily affected by HCN, resulting in large spectral features between 2 and 15 μm. Conversely, the thicker atmospheric case (10 bar surface pressure) is dominated by $CH_4$ and $C_2H_4$, allowing for its differentiation from thinner atmospheres (see Fig. 6).

### 3.3 Comparison with similar exoplanets

55 Cnc e is one of the best-studied USP super-Earths. It has received a lot of attention, and many studies have already attempted to understand the possible atmospheric composition (Fischer et al. 2008; Dawson & Fabrycky 2010; Demory et al. 2011, 2016; Tsiaras et al. 2016; Angelo & Hu 2017; Hammond & Pierrehumbert 2017). The nature of a possible nitrogen-dominated atmosphere around 55 Cnc e has been the subject of numerous theoretical studies (Miguel 2018; Zilinskas et al. 2020a, b). To further support our results, we compare them with those from Zilinskas et al. (2020b). These authors, by using one-dimensional radiative transfer and chemical kinetics models, explored a variety of possible nitrogen-rich atmospheric compositions at different temperatures, ranging from 1274 to 3730 K. They took the physical parameters of 55 Cnc e as a starting point, with a radius of 1.897 $R_\oplus$, a mass of 8.59 $M_\oplus$, and a surface pressure of 1.4 bar following Angelo & Hu (2017). They found that only atmospheres with temperatures above 2000 K and C/O ≥ 1.0 are prone to thermal inversions, due to shortwave absorption of CN. In addition, they found that HCN is the dominant carbon-nitrogen carrier for $T_{eq}$ = 2200 K and below (in cases with H = 3 × $10^{-2}$), with the HCN abundance being lower in regions with inversions due to the higher thermal dissociation. In very hot atmospheres where HCN can easily form, they also predicted that CN would show high abundances in low-pressure atmospheric regions and could have strong absorption features. Finally, as for the emission spectrum, they showed that the atmospheric features are mostly from HCN with CN having minor ones below 2 μm for cases with $T_{eq}$ = 1274 and 2200 K with hydrogen present.

Because our HD-213885b model and the benchmark 55 Cnc e are so remarkably similar in terms of radius, mass, and stellar irradiation (Espinoza et al. 2019), the computed thermal profiles, chemical composition, and emission spectra for this planet at 1 bar surface pressure are in good agreement with the Zilinskas et al. (2020b) findings. Additionally, our calculations demonstrate that disequilibrium models in very hot nitrogen atmospheres ($T_{eq}$ = 2130) are not surface pressure sensitive (Fig. 5); hence, our results for all other surface pressure scenarios are consistent with Zilinskas et al. (2020b) findings. Note that the initial atmospheric elemental compositions used in our study are similar to those of Miguel (2018) and Zilinskas et al. (2020a, b) with a C/O ≫ 1 and a hydrogen mass fraction of 3 × $10^{-2}$.

In contrast to the Zilinskas et al. (2020b) calculations, the thermal profile of our GJ-1252b model shows an inversion for $P < 10^{-5}$ bar. Our tests show that the inversion also occurs if we use the PHOENIX spectrum for the host star GJ-1252. The inversion is caused by strong atmospheric absorption by CN molecules in the near-IR together with the cool host star emitting most strongly in the near-IR. We also tested different orbital distances, thus different temperatures for the GJ-1252b model, and found that GJ-1252b-like planets with $T_{eq} \geq$ 990 K can also exhibit inversions in their atmospheres. Despite this, our emission spectrum for the 1 bar surface pressure model is consistent with the calculations of Zilinskas et al. (2020b) for $T_{eq}$ = 1274 K since most of the features are from HCN. This result is due to the fact that the upper regions where inversion occurs ($P < 10^{-5}$ bar) do not contribute much opacity to the thermal spectrum.

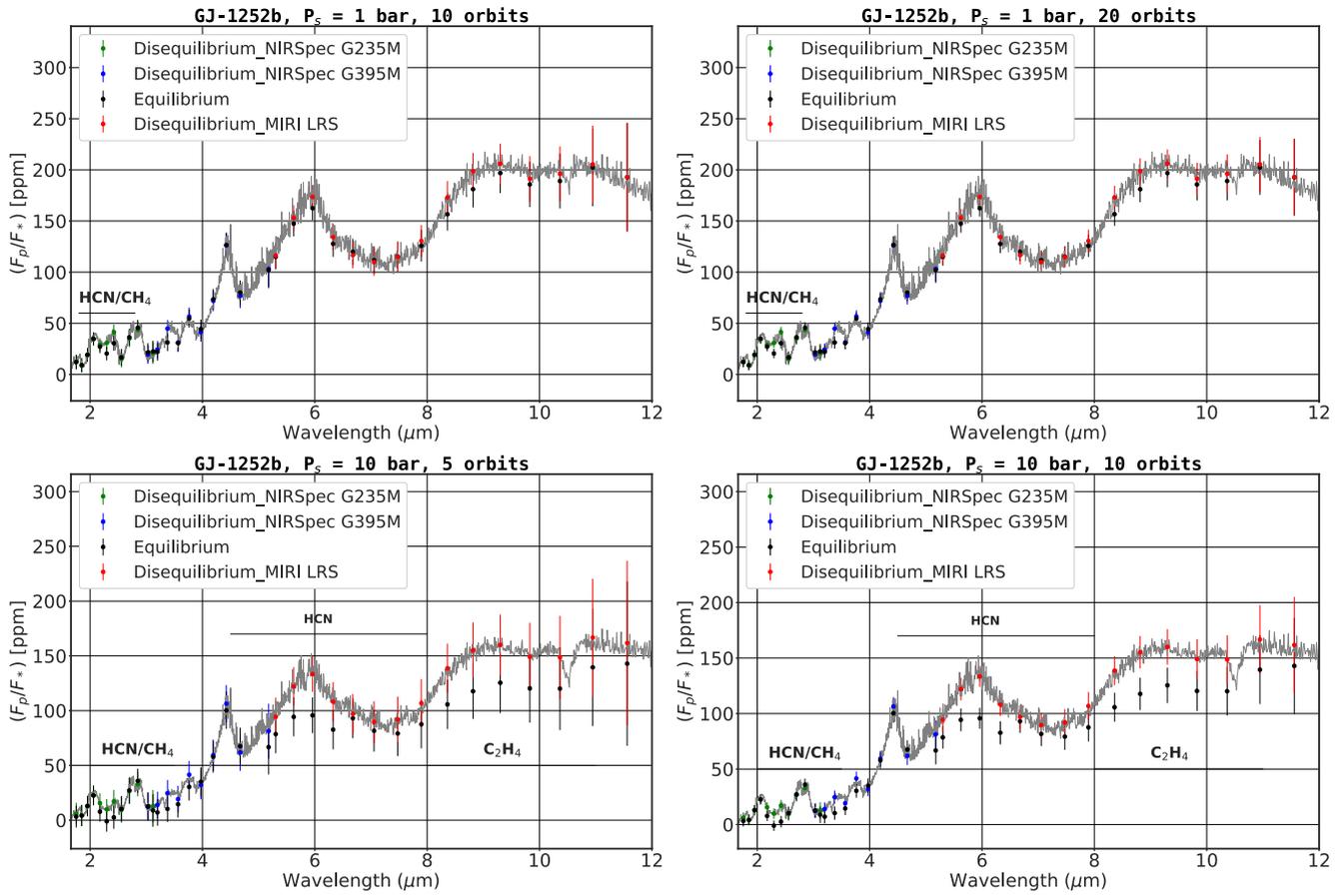

**Figure 7.** Simulated emission spectra of GJ-1252b in the 1 and 10 bar surface pressure scenarios using NIRSpec and MIRI modes. Modelled for a number of eclipse events indicated in the titles of the figures. The black error bars are for the equilibrium model and the other coloured error bars are for the disequilibrium model. The disequilibrium model spectra are underplotted in light grey. The emission spectra have been binned to resolution $R = 10$.

As previously shown in Moses et al. (2011), Venot et al. (2012), Moses (2014), and Madhusudhan et al. (2016), a lower equilibrium temperature leads to a greater divergence from chemical equilibrium. Our results confirm this as well. Overall, we see more divergence from equilibrium in our LP791-18b model than in our GJ-1252b model. This is because the lower the ambient temperature and pressure, the slower the rate at which collisional reactions will take place, thus causing the atmosphere to be farther away from equilibrium.

## 4 CONSTRAINING DISEQUILIBRIUM EFFECT WITH *JWST*

In this section, we aim to investigate whether the difference between equilibrium and disequilibrium chemistry can be observed in the atmospheres of USP super-Earth planets with the *JWST*. We previously showed that for HD-213885b, the strongest deviations from equilibrium can only be seen in the 10 bar surface pressure scenario, particularly at 13.9 μm due to the presence of HCN. Similarly, in the 0.1 bar surface pressure scenario of GJ-1252b, differences were seen at 13.9 μm and near 20 μm due to HCN and HCN/CH$_4$ features, respectively. These characteristics could potentially be observed with the MIRI/*JWST* instrument (channels 3 and 4, wavelength range 11.52 and 28.3 μm) using the medium-resolution spectrometer (MIRI MRS). MIRI photometry may also be able to target the longer wavelengths, and consequently, these features. However, we could not calculate the expected noise level and required observing time for MIRI MRS observing mode because the current version of PANDEXO does not support it. Nevertheless, we expect these relatively strong features to be detectable with *JWST*.

Here, we discuss the three surface pressure scenarios for the planet LP791-18b as well as the 1 and 10 bar surface pressure cases for the planet GJ-1252b, because this is where the effects of chemical disequilibrium are most pronounced, particularly in the 2–11 μm wavelength range covered by *JWST*'s instruments. To investigate this, we use the *JWST* instrument noise simulator PANDEXO[13] (Batalha et al. 2017) to simulate secondary eclipse observations of the planets GJ-1252b and LP791-18b. These planets are bright enough to offer high S/N while faint enough to be observable with all of *JWST*'s instruments. We therefore choose to observe each of our planets with NIRSpec/G235M from 1.66 to 3 μm ($R = 1000$), NIRSpec/G395M mode from 2.9 to 5 μm ($R = 1000$), and MIRI/LRS mode from 5 to 11 μm ($R = 100$). NIRSpec Prism would saturate on GJ-1252b and was therefore not considered in this analysis. These observing modes were chosen because they offer nearly full wavelength coverage from 2 to 11 μm, where the disequilibrium features can be distinguished from equilibrium ones.

PANDEXO uses the Phoenix Stellar Atlas models (Husser et al. 2013) and requires a noise floor and saturation limit that are instrument dependent. We set the saturation level to 80 per cent full well and the total observation time to two times the eclipse duration. In line with previous studies (Beichman et al. 2014; Ferruit et al. 2014; Greene et al. 2016; Rocchetto et al. 2016; Batalha et al. 2017; Chouqar et al. 2020; Fortenbach & Dressing 2020), we assume the noise floor to be 25 and 50 ppm for the NIRSpec and MIRI modes, respectively. After testing various binning resolutions, we find that the models are distinguishable for a resolution of $R = 10$. We follow the same approach as of Shulyak et al. (2020), and we estimate the detectability of spectroscopic features by using the Chi-square test between the noise-free spectrum of equilibrium and disequilibrium models and simulated disequilibrium observations, respectively. These $\chi^2$ values were computed within spectral regions affected by disequilibrium processes.

---
[13] https://natashabatalha.github.io/PandExo

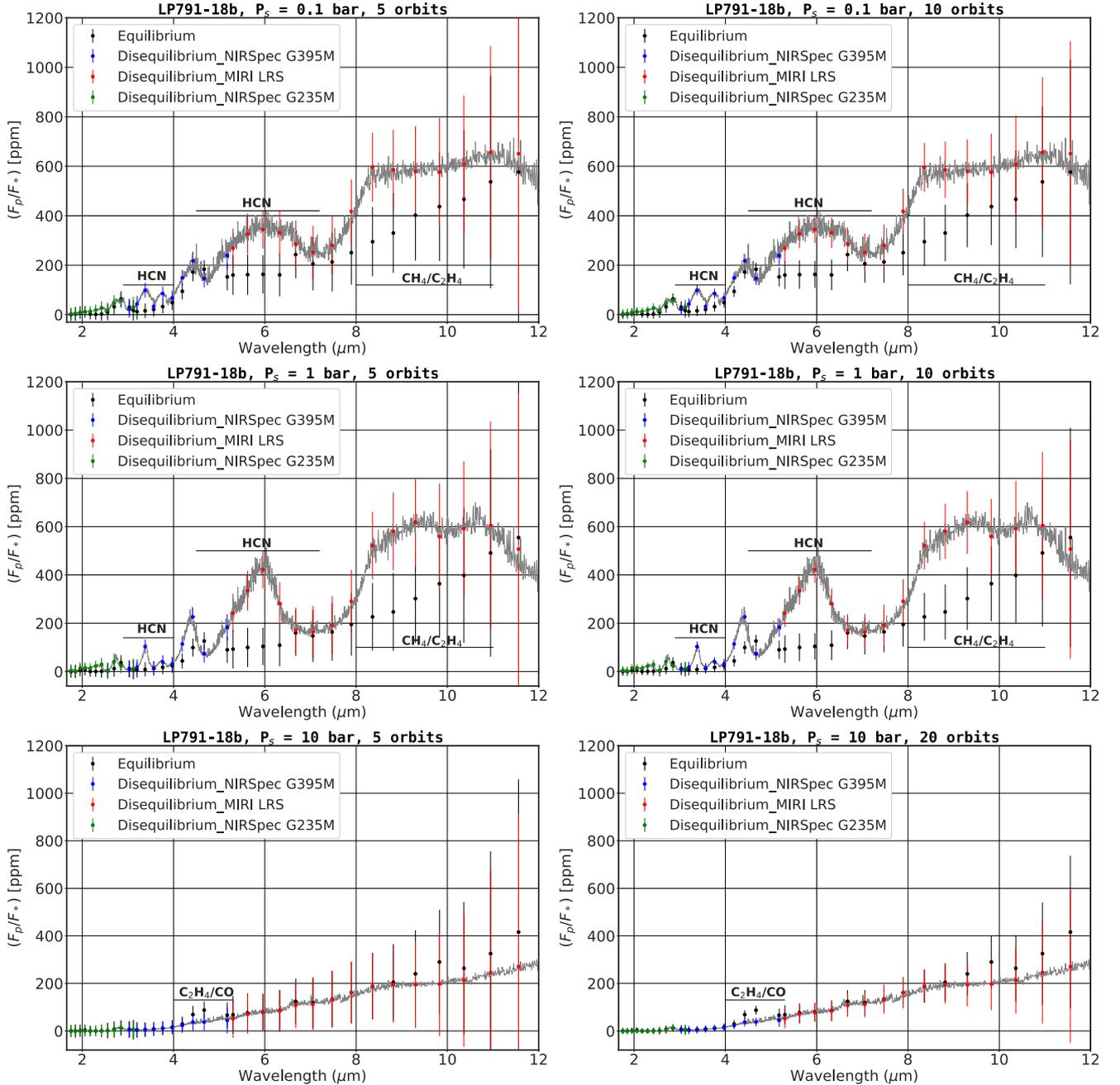

**Figure 8.** Same as on Fig. 7, but for the planet LP791-18b in all the surface pressure scenarios considered in this work.

Figs 7 and 8 show our synthetic *JWST* emission spectra with error bars representing the number of observed secondary eclipses indicated in the titles of the figures for both GJ-1252b and LP791-18b in different surface pressure scenarios. As expected, the more observations, the clearer the separation between equilibrium and disequilibrium. For GJ-1252b (Fig. 7), the divergence caused by HCN/CH$_4$ and HCN signatures between 3–4 and 5–7 µm can be seen with 10 eclipses using NIRSpec/G395M and MIRI/LRS modes in the 10 bar surface case, respectively. The deviation due to the HCN/CH$_4$ features between 2 and 3 µm would require 20 and 10 eclipse events to be observed with the NIRSpec/G235M mode in the 1 and 10 bar surface scenarios, respectively. For LP791-18b, our estimates show that the effect of chemical disequilibrium would need five observed eclipses, using the HCN feature at around 3.3 μm, to be detected above the $3\sigma$ threshold in the 0.1 bar surface case, while 10 eclipses would be needed to detect the effect caused by HCN at around 5–7 μm. In the 1 bar surface case, we can see the divergence caused by HCN features with only five eclipses. The noise is large at wavelengths $\lambda > 10$ μm, making it difficult to distinguish between equilibrium and disequilibrium using the $C_2H_4$ and $CH_4/C_2H_4$ features at around 10 μm for both GJ-1252b and LP791-18b, respectively. This large noise is due to a dramatic drop in the MIRI-LRS throughput (Shulyak et al. 2020). The separation caused by $C_2H_4$/CO at around 4.3 μm in the 10 bar surface scenario would require about 20 eclipses to be detectable. Overall, both NIRSpec and MIRI instruments can make it possible to detect differences caused by disequilibrium chemistry via eclipse spectroscopy by targeting HCN in the case of thinner atmospheres ($P_s \leq 1$ bar) and $C_2H_4$/CO in the case of thicker ones for cooler planets (for LP791-18b-like planets) and HCN/$CH_4$ in the case of thicker atmospheres ($P_s \geq 1$) for warmer planets (for GJ-1252b-like planets). Additionally, our grids show that these surface pressure scenarios can be distinguished. LP791-18b-like planets with thinner atmospheres ($P_s \leq 1$ bar) are dominated by the photochemically produced HCN, shaping their emission spectra and showing large spectral features between 2 and 15 μm, while $CH_4$ and $C_2H_4$ dominate thicker atmospheric cases ($P_s > 1$ bar), making it possible to distinguish them from thinner ones. For GJ-1252b-like planets, thinner atmospheres, with $P_s < 1$ bar, are dominated by $CH_4$ compared to the thicker ones ($P_s \geq 1$ bar) where HCN becomes the dominant spectrally active species. Looking for these species using NIRSpec/G235M and MIRI/LRS modes can give clues about the thickness of these atmospheres. Along with other species (e.g. ammonia), methane and hydrogen cyanide have been found to be most sensitive to the existence of surfaces in a hydrogen-dominated sub-Neptune atmosphere (Yu et al. 2021). On the other hand, Tsai et al. (2021a) suggested that HCN is not applicable to determine the pressure level of the surface in such atmospheres. Instead, they found ammonia to be the most unambiguous proxy to infer surfaces in the era of the *JWST*. However, compared to $H_2$-dominated atmospheres, the formation pathways for nitrogen-bearing species in cool $N_2$-dominated atmospheres tend to favour the production of HCN. Ammonia is dissociated into its constituent atoms, giving rise to nitrogen gas, which is the most abundant species in our atmospheres. Additionally, the production of HCN in our cooler planet is enhanced by the coupled $NH_3$–$CH_4$ photochemistry, contributing also to the depletion of ammonia. Our models show that the detection of HCN in cooler planets can facilitate the determination of shallow-surface scenarios using *JWST*. For hotter planets, such as HD-213885b, we see that it would be difficult to differentiate between various surface pressure scenarios since the higher temperature of these atmospheres keeps CN and HCN the dominant species.

## 5 CONCLUSION

In our study, we presented a grid of radiative–convective temperature profiles, photochemical mixing ratios, and emission spectra for various surface pressure scenarios ranging from $10^{-1}$ to 10 bar, on cool (LP791-18b), warm (GJ-1252b), and hot (HD-213885b) USP super-Earths. We explored how various surface pressures affect thermal profiles. We also performed a case study comparing the atmospheric abundances and the dayside emission spectra of these planets, in and out of chemical equilibrium at the different surface pressure scenarios in order to explore the observability of differences due to disequilibrium processes with *JWST*. Our grids demonstrate that various surface levels can significantly affect the thermal profiles, and alter the atmospheric abundances, and thus, the emission spectra of a nitrogen-dominated USP super-Earth atmosphere. We found that along with hotter USP rocky planets, warmer planets with $T_{eq} \geq 990$ K orbiting M dwarfs may also show inversions in a nitrogen-dominated atmosphere. Modelling atmospheres of USP super-Earths orbiting M dwarfs requires additional attention in future work. We found that in addition to disequilibrium processes, HCN, $CH_4$, and $C_2H_4$ abundances are also sensitive to different surface pressures, making it possible to distinguish equilibrium from disequilibrium models, particularly for lower surface levels in cooler planets and higher surface pressures in warmer planets. These species can help distinguish the existence and role of photochemical reactions. Furthermore, these species can be used to differentiate between the different surface levels, giving clues about the thickness of a nitrogen-dominated USP super-Earth atmosphere. We showed that both NIRSpec and MIRI instruments are viable for the purpose of distinguishing disequilibrium chemistry from equilibrium chemistry via eclipse spectroscopy on cooler planets (LP791-18b-like planets) and on warmer planets with thicker atmospheres ($P_s \geq 1$ bar for GJ-1252b-like planets). With these instruments, we cannot readily observe the contribution from disequilibrium chemistry in HD 213885b-like planets given our inputs because it is too hot. At such high temperatures, thermochemical equilibrium dominates over disequilibrium processes, making it difficult to distinguish them.

## ACKNOWLEDGEMENTS

The authors wish to thank Matej Malik for his help with the HELIOS code. We also thank the anonymous referee for the illuminating comments. We acknowledge the support offered by computational resources of HPC-MARWAN (hpc.marwan.ma) provided by the National Center for Scientific and Technical Research (CNRST), Rabat, Morocco.

## DATA AVAILABILITY

The numerical results underlying this article were calculated using the open-source codes HELIOS,[14] FASTCHEM,[15] VULCAN,[16] and PETITRADTRANS.[17] The opacity data were obtained from a public data base.[18] Other relevant data will be shared on reasonable request to the corresponding author.

## REFERENCES

Angelo I., Hu R., 2017, AJ, 154, 232
Banks P., Kockarts G., 1973, Aeronomy, Partie 2. Acad. Press, USA
Batalha N. et al., 2011, ApJ, 729, 27
Batalha N. E. et al., 2017, PASP, 129, 064501
Beichman C. et al., 2014, PASP, 126, 1134
Bourrier V. et al., 2018, A&A, 619, A1
Burrows A., Ibgui L., Hubeny I., 2008, ApJ, 682, 1277
Chen J., Kipping D., 2016, ApJ, 834, 17

---

[14] https://github.com/exoclime/HELIOS
[15] https://github.com/exoclime/FastChem
[16] https://github.com/exoclime/VULCAN
[17] http://gitlab.com/mauricemolli/petitRADTRANS
[18] http://www.exomol.com/

Chouqar J., Benkhaldoun Z., Jabiri A., Lustig-Yaeger J., Soubkiou A., Szentgyorgyi A., 2020, MNRAS, 495, 962
Chubb K. L. et al., 2020, A&A, 646, A21
Crossfield I. J. M. et al., 2019, ApJ, 883, L16
Crossfield I. J. et al., 2022, ApJ, 937, L17
Dai F. et al., 2017, AJ, 154, 226
Dawson R. I., Fabrycky D. C., 2010, ApJ, 722, 937
Demory B. O. et al., 2011, A&A, 533, A114
Demory B.-O., Gillon M., Seager S., Benneke B., Deming D., Jackson B., 2012, ApJ, 751, L28
Demory B.-O. et al., 2016, Nature, 532, 207
Drummond B., Tremblin P., Baraffe I., Amundsen D. S., Mayne N. J., Venot O., Goyal J., 2016, A&A, 594, A69
Espinoza N. et al., 2019, MNRAS, 491, 2982
Esteves L. J., de Mooij E. J. W., Jayawardhana R., Watson C., de Kok R., 2017, AJ, 153, 268
Ferruit P., Birkmann S., Böker T., Sirianni M., Giardino G., de Marchi G., Alves de Oliveira C., Dorner B., 2014, in Oschmann J. M., J., Clampin M., Fazio G. G., MacEwen H. A., eds, Proc. SPIE Conf. Ser. Vol. 9143, Space Telescopes and Instrumentation 2014: Optical, Infrared, and Millimeter Wave. SPIE, Bellingham, p. 91430A
Fischer D. A. et al., 2008, ApJ, 675, 790
Fortenbach C. D., Dressing C. D., 2020, PASP, 132, 054501
Greene T. P., Line M. R., Montero C., Fortney J. J., Lustig-Yaeger J., Luther K., 2016, ApJ, 817, 17
Grimm S. L., Heng K., 2015, ApJ, 808, 182
Hammond M., Pierrehumbert R. T., 2017, ApJ, 849, 152
Howard A. W. et al., 2013, Nature, 503, 381
Husser T. O., Wende-von Berg S., Dreizler S., Homeier D., Reiners A., Barman T., Hauschildt P. H., 2013, A&A, 553, A6
Koll D. D., 2022, ApJ, 924, 134
Kreidberg L. et al., 2019, Nature, 573, 87
Madhusudhan N., Agúndez M., Moses J. I., Hu Y., 2016, Space Sci. Rev., 205, 285
Malik M. et al., 2017, AJ, 153, 56
Malik M., Kempton E. M.-R., Koll D. D., Mansfield M., Bean J. L., Kite E., 2019, ApJ, 886, 142
Miguel Y., 2018, MNRAS, 482, 2893
Miguel Y., Kaltenegger L., Fegley B., Schaefer L., 2011, ApJ, 742, L19
Miller-Ricci E., Fortney J. J., 2010, ApJ, 716, L74
Mollière P., Wardenier J. P., van Boekel R., Henning Th., Molaverdikhani K., Snellen I. A. G., 2019, A&A, 627, A67
Moses J. I., 2014, Phil. Trans. R. Soc. A: Math. Phys. Eng. Sci., 372, 20130073
Moses J. I., Bézard B., Lellouch E., Gladstone G. R., Feuchtgruber H., Allen M., 2000, Icarus, 143, 244
Moses J. I., Visscher C., Keane T. C., Sperier A., 2010, Faraday Discuss., 147, 103
Moses J. I. et al., 2011, ApJ, 737, 15
Ofir A., Dreizler S., 2013, A&A, 555, A58
Pepe F. et al., 2013, Nature, 503, 377
Queloz D. et al., 2009, A&A, 506, 303
Rappaport S., Sanchis-Ojeda R., Rogers L., Levine A., Winn J., 2013, ApJ, 773, L15
Ricker G., 2015, J. Astron. Telesc. Instrum. Syst., 1, 014003
Ridden-Harper A. R. et al., 2016, A&A, 593, A129
Rocchetto M., Waldmann I. P., Venot O., Lagage P.-O., Tinetti G., 2016, ApJ, 833, 120
Rogers J. C., Apai D., López-Morales M., Sing D. K., Burrows A., 2009, ApJ, 707, 1707
Rouan D., Deeg H. J., Demangeon O., Samuel B., Cavarroc C., Fegley B., Léger A., 2011, ApJ, 741, L30
Rowe J. F. et al., 2006, ApJ, 646, 1241
Rugheimer S., Kaltenegger L., Segura A., Linsky J., Mohanty S., 2015, ApJ, 809, 57
Sanchis-Ojeda R., Rappaport S., Winn J. N., Kotson M. C., Levine A., Mellah I. E., 2014, ApJ, 787, 47
Santerne A. et al., 2018, Nat. Astron., 2, 393
Schaefer L., Fegley B., 2009, ApJ, 703, L113
Schaefer L., Lodders K., Fegley B., 2012, ApJ, 755, 41
Shporer A. et al., 2020, ApJ, 890, L7
Shulyak D., Lara L. M., Rengel M., Nèmec N.-E., 2020, A&A, 639, A48
Smith A. M. S. et al., 2017, MNRAS, 474, 5523
Stock J. W., Kitzmann D., Patzer A. B. C., Sedlmayr E., 2018, MNRAS, 479, 865
Sudarsky D., Burrows A., Pinto P., 2000, ApJ, 538, 885
Tsai S.-M., Lyons J. R., Grosheintz L., Rimmer P. B., Kitzmann D., Heng K., 2017, ApJS, 228, 20
Tsai S.-M., Innes H., Lichtenberg T., Taylor J., Malik M., Chubb K., Pierrehumbert R., 2021a, ApJ, 922, L27
Tsai S.-M., Malik M., Kitzmann D., Lyons J. R., Fateev A., Lee E., Heng K., 2021b, ApJ, 923, 264
Tsiaras A. et al., 2016, ApJ, 820, 99
Venot O., Hébrard E., Agúndez M., Dobrijevic M., Selsis F., Hersant F., Iro N., Bounaceur R., 2012, A&A, 546, A43
Whittaker E. A. et al., 2022, AJ, 164, 258
Winn J. N., Sanchis-Ojeda R., Rappaport S., 2018, New Astron. Rev., 83, 37
Yu X., Moses J. I., Fortney J. J., Zhang X., 2021, ApJ, 914, 38
Zeng L., Sasselov D. D., Jacobsen S. B., 2016, ApJ, 819, 127
Zieba S. et al., 2022, A&A, 664, A79
Zilinskas M., Miguel Y., Mollière P., min Tsai S., 2020a, MNRAS, 494, 1490
Zilinskas M., Miguel Y., Lyu Y., Bax M., 2020b, MNRAS, 500, 2197

This paper has been typeset from a T<sub>E</sub>X/LaT<sub>E</sub>X file prepared by the author.